# Hierarchical representation of socio-economic complex systems according to minimal spanning trees

Andrzej JARYNOWSKI[*], Andrzej BUDA[†]

**Abstract:** We investigate hierarchical structure in various complex systems according to Minimum Spanning Tree methods Firstly, we investigate stock markets where the graph is obtained from the matrix of correlations coefficient computed between all pairs of assets by considering the synchronous time evolution of the difference of the logarithm of daily stock price. The hierarchical tree provides information useful to investigate the number and nature of economic factors that have associated a meaningful economic taxonomy. We extend this method on other financial market – money exchange (FOREX) and commodity - phonographic market (where we have artists instead of stocks) and get information on which music genre is meaningful according to customers. We continue to use this method in social systems (sport, political parties and pharmacy) to investigate collective effects and detect how single element of the system influences on the other ones. The level of correlations and Minimum Spanning Trees in various complex systems is also discussed.

## 1. Introduction to hierarchical representation

Mathematical methods have become more and more popular and successfully applied in explanations of phenomena observed in real world social, economic and biological systems. We propose building a meaning-full representation to show complex relation between agencies in various systems (Green and Bossomaier 2000). We believe these visualization methods and its quantitative results can be exploited in research on markets and other social systems (not only in examples we provided). In contrast to many other quantitative methods like statistical regressions and data-ming procedures, hierarchical representation has relatively transparent structure. Within the project, we also design a novelty life-time approach to see change in hierarchical structure in time. In literature, there are attempts to model financial and commodity markets, but they do not succeed in social sciences and medicine. It is worth to see that relations discovered by our methodology, often have not been satisfactorily understood yet. Thus, we expect that our research project will be a huge step for better understanding of the processes and rules of the social systems evolution (Watts and Strogatz 1998). Because this type of relationship has not been sufficiently explored yet.

---

[*] Smoluchowski Institute, Jagiellonian University in Cracow
[†] Institute of Nuclear Physics in Cracow, Polish Academy of Science

Recently, the knowledge of complex system tools for economy, sociology and medicine such as networks and hierarchical representation has undergone an accelerating growth, however all models of such system are incomplete without real data, especially register-based (Mezard 1987). Complex systems are natural or social systems that consist of a large number of nonlinear interacting elements. The requirement to understand phenomenon encourages cooperation between various registering institutions, which, in turn, exerts a pressure on collecting data for simple analysis by many researchers who work on new models and use complex tools often taken from other disciplines. The most exciting property of these systems is the existence of emergent phenomena which cannot be simply derived or predicted solely from the knowledge of the system' structure and the interactions between their individual elements. However, physics methodology proves helpful in many issues of complex systems properties including the collective effects and their coexistence with noise, long range interactions, the interplay between determinism and flexibility in evolution, scale invariance, criticality, multifractality (Oswiecimka, Kwapien et al. 2011) and hierarchical structure (Grabowski and Kosiński 2004). The aim of our article is to fill in the gap between social and medical science analyzes with complex system approach (Kwapień 2012) by applying meaningful taxonomy developed previously in field of applied mathematics, complex systems and computer science. In this paper we investigate various complex systems like financial (stock exchange: DJIA, DAX, FTSE1000, WIG20, money exchange: FOREX) and commodity (phonographic) market, social systems of political parties, sport (Polish Football League) and pharmacy, but from the hierarchical structure point of view.

Tab. 1. Presented example data structure

| Data type | Time span | No assets | Signal | Length of series | Asset category |
|---|---|---|---|---|---|
| Stock exchange | 1997-2008 | ~ 30 | Prize | ~1000 | Stock company |
| Money exchange | 2002-2013 | ~30 | Relative exchange rate | ~1000 | Currency |
| Phonographic market | 2004- 2014 | 30 | Record sale | ~400 | Artists |
| Politics | 2003-2014 | ~ 10 | % of support in polls | ~100 | Political party |
| Football | 2003-2004 | ~20 | Points in game | ~40 | Team |
| Pharmacy | NA~2004 | ~10 | Points in survey | ~50 | Health indicator |

## 2. The correlation and its interpretation

Initially, we analyze correlations matrices of signal (Mantegna 1999). The correlation coefficient defines a degree of similarity between the synchronous time evolution of a pair of assets, where we took of underlying value (prize, sale, points, preference, etc.). There are many measures of correlation like mutual information or Manhattan, but we choose the simplest one is linear (Pearson).

$$\rho_{ij} = \frac{<Y_iY_j>-<Y_i><Y_j>}{\sqrt{(<Y_i^2>-<Y_i>^2)(<Y_j^2>-<Y_j>^2)}} \quad (1)$$

where $i$ and $j$ are the numerical labels of assets, $Y_i$ is the return or signal (underlying).

Definition of return: $Y_i = \ln[P_i(t)] - \ln[P_i(t-1)]$ where $P_i(t)$ is the signal $i$ at time $t$. The statistical average is a temporal average performed on all the trading days of investigated time period. By definition, $\rho_{ij}$ may vary from -1 to 1. The matrix of correlation coefficients is a symmetric matrix with $\rho_{ii}$ and the $n(n-1)/2$ correlation coefficients characterize the matrix completely. Every correlation from that matrix based on two vectors containing $P_i$ and $P_j$: the time series of signal $i$ and $j$ for every given time interval. The correlation coefficient reflects similarity between assets. It can be used in building the hierarchical structure in system and finding the taxonomy that allows isolating groups of assets.

Three levels of correlations can be introduced:
1. Strong (strongly correlated pair of assets)      $\rho \in [1/2, 1]$ ;
2. Weak (weakly correlated pair of assets)         $\rho \in [0, 1/2)$ ;
3. Negative (anti-correlated pair of assets)       $\rho \in [-1, 0)$ .

Tab. 2. Number of strongly, weakly and negatively correlated pairs in portfolios.

| Correlated pairs | strongly | weakly | Negatively |
|---|---|---|---|
| DJIA | 9 | 426 | 0 |
| DAX | 205 | 119 | 1 |
| WIG 20 | 1 | 188 | 1 |
| Football | 2 | 60 | 58 |
| Politics | 0 | 7 | 8 |
| Phonographic market | 3 | 72 | 375 |

## 3. Minimal Spanning Tree and hierarchical diagrams

The correlations matrix can also be used in order to classify the artists into clusters. The distance between assets is defined by:

$$d_{ij} = \sqrt{2(1-\rho_{ij})} \quad (2)$$

and is associated with correlation coefficients. With this choice, $d_{ij}$ fulfills three axioms of an Euclidean metric:

$d_{ij} = 0$ if and only if $i = j$,

$$d_{ij} = d_{ji},$$

$$d_{ij} < d_{ik} + d_{kj}$$

For that, we could build the Minimal Spanning Tree (MST) and Hierarchical diagrams for the portfolio of assets. Let illustrate it on financial market – DJIA [Fig. 1C]. Firstly we detect link d for the strongest correlation (the shortest distance). For example, C-JPM is the strongest correlation (0.72) from the DJIA portfolio. Thus, we start to build the Minimal Spanning Tree from the distance $d_{C\text{-}JPM} = 0.75$. The second strongest correlation (0.68) is AXP-C and we could add this additional link (d = 0.8) to C. The third strongest correlation is AXP-JPM (0.65), but AXP and JPM is already connected by C. Another strongest correlation is GE-AXP (0.61) so we add this link to AXP. After joining all the 30 stocks we have the complete Minimal Spanning Tree (*n*-1 links) for the DJIA portfolio. It reflects sectors and subsectors of the economy in the investigated time period (according to daily stocks closing price history only). We present also other markets

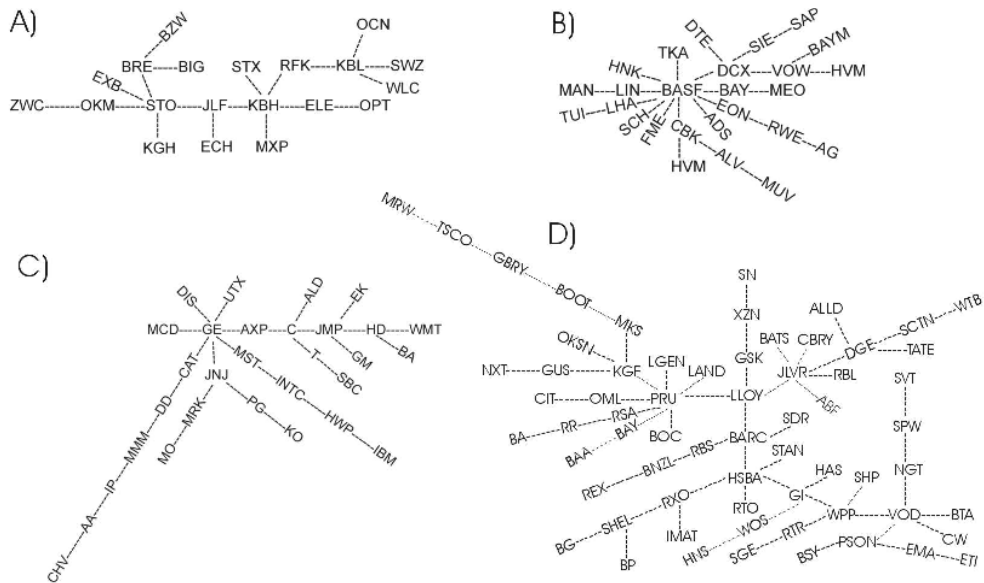

Fig. 1. Minimum Spanning Tree (MST) connecting the stocks used to compute (A. Buda 2012): A) Warszawski Indeks Giełdowy (WIG 20); B) Deutsche Aktienindex (DAX); C) Dow Jones Industrial Average (DJIA); D) FTSE 1000

The Minimum Spanning Trees can be also obtained in Foreign Exchange market (FOREX). However, the structure of these trees might depend on reference frame, because all values $P_i(t)$ have to be expressed by the basic currency (Fig. 2). We investigated 38 currencies (including gold) in the investigated period and detected geographical dependences between currencies.

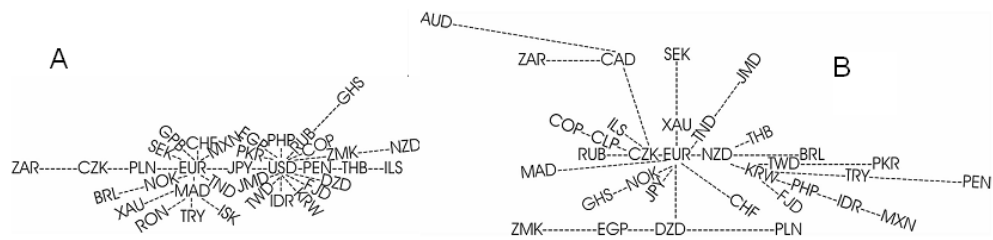

Fig. 2. The Minimum Spanning Tree obtained for 37 currencies (2002-2012). The basic currency is A) USD, B) AUD.

Another example of The Minimal Spanning Tree (MST) comes from commodity markets, like the global phonographic market, where artists are the assets (Fig. 3). Although the hierarchical structure in financial markets reflects the Forbes classifications of stocks in the industry sectors and sub-sectors (Fig. 1), the analysis of the MST and correlations between artists does not always fit to music genres classified by the Billboard and other music magazines (Ordanini 2006). The Minimal Spanning Tree reveals sectors that belong to rap (Kanye West, 50 Cent, Usher), rock (Green Day, Kings Of Leon, The Beatles, U2, Evanescence or Coldplay, Nickelback, Kelly Clarkson), soul and r'n'b (Adele, Britney Spears, Alicia Keys, Beyonce, Michael Jackson), but according to customers, there is no sector for pure pop music (Buda and Jarynowski 2013).

Instead of pop, in the middle of the MST we have a celebrity sector that contain Lady Gaga, Rihanna, Bruce Springsteen, Pink, Jay-Z, The Beatles, Black Eyed Peas and Justin Timberlake. Although they represent various styles and genres, the only common thing they have is fame, high record sales, popularity. Most of them were popular before 2003.

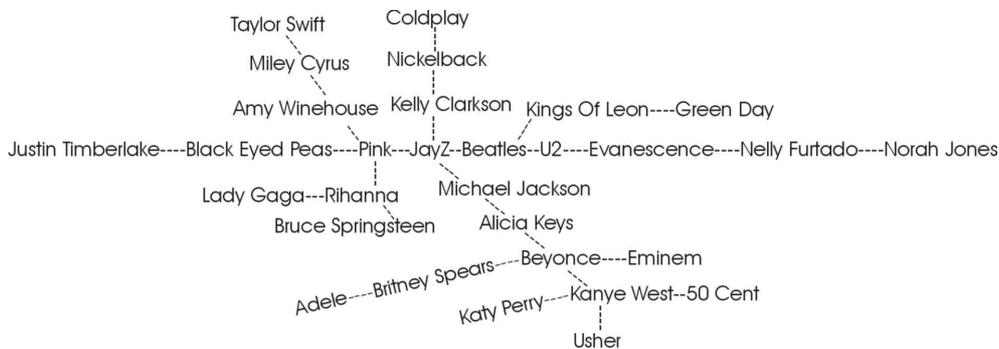

Fig. 3. The Minimum Spanning Tree obtained for the 30 most popular artists in global phonographic market (2003-2011).

One of the most significant events in Poland after communism broke up was the emergence of internal conflict after Polish president's plane crash 10.04.2010. We investigate relations between political parties in Poland during last decade. We provide analysis of preferences of voters (via polls), personal transfers between parties and public opinion associations before and after Smolensk movement (Airport in Russia where the catastrophe had happened), that represent medial perception of distances between parties. We observe a topological difference in Minimal Spanning Trees representing connections between political parties - before and after critical phenomenon (Fig. 4). For example, polarization occurs on PO-PiS line (these parties are no longer neighbors in hierarchical diagram and the Minimal Spanning Tree, because of strong negative correlations according to opinion polls after 10.04.2010). Before Smolensk plane crash every link in the MST had representation in transitions between political parties, because surprisingly people changed political parties directly according to the MST links.

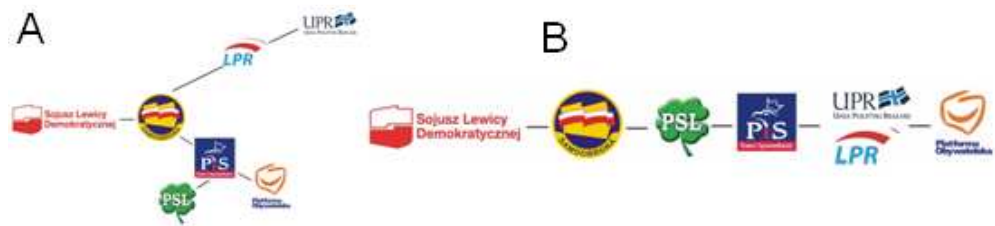

fig. 4. The Minimum Spanning Tree obtained for Polish political parties according to opinion polls A) 2006-2010 (before Smolensk), B) 2006-2013 (after Smolensk).

The Minimal Spanning Trees can also describe social systems without given time series. For the presentation, we apply non-linear techniques like the quality of life assessment analyze made in Poznań on 86 patients using a questionnaire (WHOQOL - BREF) consisting of 26 questions (Najmrodzka, 2012). We focused on four areas: somatic (physical), psychological, social and environmental impacts, and additional questions regarding the status of self- subjective perception of health and quality of life. Results were normalized to calculate the correlation coefficients between investigated traits (assets). We found that the traditional medical methods of testing hypotheses receive only a few statistically significant correlations. It is a triangle: sex, weight and height, and health status coupled with quality of life. Traditional ways of presentation the characteristics that affect the quality of life are unsatisfactory. On the other hand, visualization by MST gives clarity of diagnosis: the elements that affect the quality of life directly and indirectly (Fig. 5). Thus, the MST gives new and relevant quality for epidemiological studies.

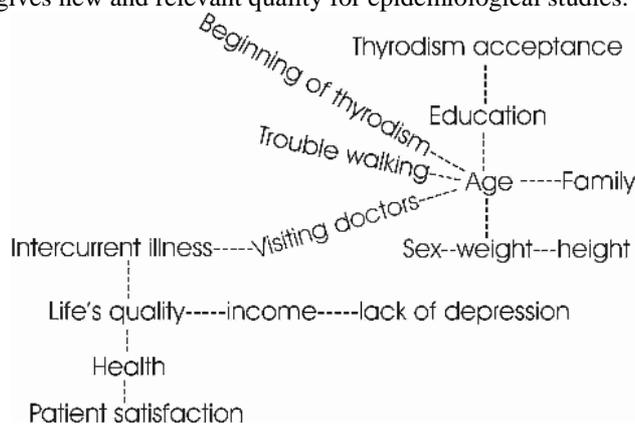

Fig. 5. The Minimum Spanning Tree obtained for the features of the people who suffer from thyroid.

Subdominant ultrametric space that provides hierarchy of correlations in other perspective like diagrams. This method describes collective relations between elements like teams in football league while in few seasons of Polish League corruption had a big role (Jarynowski 2010). In recent years prosecution in Poland has been investigating several clubs, referees and players because of corruption procedures. We study the statistical properties of results in Polish League, looking for evidence of non-sport activity. We treat league as a complex system and we use tools from statistical physics to research some of its properties by investigating system of ranked elements (Fig. 6) in time series and finding which clubs play for another profits and analyzing statistical situation before and after matches, which were stated by the court as those in which a crime has been committed. This research is dedicated to release Polish Football from problem of corruption.

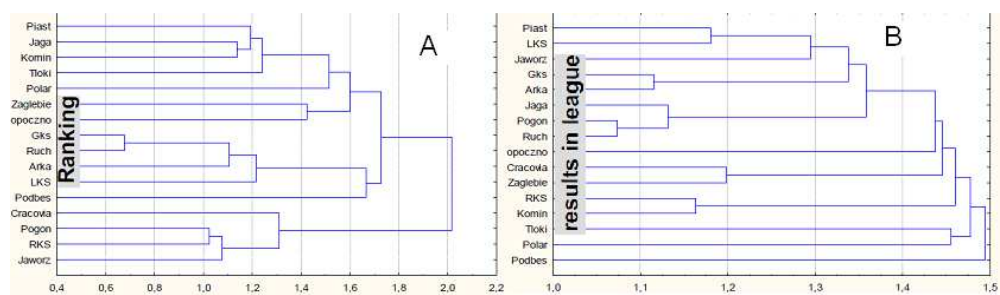

Fig. 6. Hierarchical representation of First Polish League season 2003/2004. A) Signal is ranking (place in table), B) signal is result in game.

## 4. Conclusions

Results of our visualization analysis may develop new (more reliable) dynamic description of complex systems. Methodology of MST diagrams based on correlation coefficients in social and economic phenomena makes data more understandable. It has been already shown in financial markets where it is possible to isolate groups of stocks or currencies that make sense from economic point of view. Our aim is to make mathematical, quantitative perspective, because experts in social science used to work with unsatisfactory methods. This is necessary for better understanding of a investigated system. It is also possible to obtain more reliable comparison between systems (as we presented in this paper). Multiscale dimension mapping methodology allows us to present relation in both local and global scale. These methods allow us in particular focus on new scientific problems. Because the MST methods have given a better description of people who suffer from thyroid. Thus, it is possible to isolate previously unknown chains of influences in the Minmal Spanning Tree that have direct or indirect impact on the quality of life.

In political studies, we ask a questions: Is polarization observed in both regimes or only in the media perception? And what does characteristic fingerprints of such a critical event to political ties in general (in meaningful commodity market taxonomy where 'prize' of political party is en equivalent to opinion polls). While social scientists try to understand mechanisms of political polarization with underlying psychological and political patterns, our hierarchical network analysis with application of natural language processing and text mining, are great supplementary tools for that. The aim of this preliminary exploratory quantitative study was to generate questions and hypotheses, which could be carefully follow by qualities methods in the future.

In phonographic market the conclusion is, that from economical point of view, pop music does not exist in the way that record companies think (superstar cluster instead of genre like). Pop music is a term that originally derives from an abbreviation of "popular" and its extension to music genre is not allowed. We successfully utilized both a qualitative approach (aimed at the so-called conscious consumers) and a quantitative one (with methods of statistical physics, aimed at mass consumers only weakly related to the phonographic and music industry).

Finally, we expect that our research will have contribution in the knowledge of the dynamic phenomena occurring in complex systems especially those of social and economic nature.